\documentclass[twocolumn,prb,amsmath,amssymb]{revtex4}

\usepackage{graphics}
\usepackage{graphicx}
\usepackage{dcolumn}
\usepackage{bm}
\usepackage{latexsym}

\def\vp{\varphi}

\def\eps{\epsilon}

\def\om{\omega}

\def\be{\begin{equation}}

\def\ee{\end{equation}}

\def\bea{\begin{eqnarray}}

\def\eea{\end{eqnarray}}

\def\bc{\begin{center}}

\def\ec{\end{center}}

\def\bs{\bigskip}

\def\ni{\noindent}

\def\ov{\over}

\def\nonum{\nonumber}

\begin{document}

\title{Tails of the dynamical structure factor \\ of 1D spinless fermions beyond the Tomonaga approximation}


\author{Sofian Teber}

\email{steber@ictp.trieste.it}

\affiliation{The Abdus Salam ICTP, Strada Costiera 11, 34014, Trieste, Italy}

\date{\today}

\begin{abstract}
We consider one-dimensional (1D) interacting spinless fermions with a non-linear spectrum in a clean quantum wire (non-linear bosonization). We compute diagrammatically the 1D dynamical structure factor, $S(\om,q)$, beyond the Tomonaga approximation focusing on it's tails, $|\om| \gg vq$, {\it i.e.} the 2-pair excitation continuum due to forward scattering. Our methodology reveals three classes of diagrams: two "chiral" classes which bring divergent contributions in the limits $\om \rightarrow \pm vq$, {\it i.e.} near the single-pair excitation continuum, and a "mixed" class (so-called Aslamasov-Larkin or Altshuler-Shklovskii type diagrams) which is crucial for the f-sum rule to be satisfied. We relate our approach to the $T=0$ ones present in the literature. We also consider the $T\not=0$ case and show that the 2-pair excitation continuum dominates the single-pair one in the range: $|q|T/k_F \ll \om \mp vq \ll T$ (substantial for $q \ll k_F$). As applications we first derive the small-momentum optical conductivity due to forward scattering: $\sigma \sim 1/\om$ for $T \ll \om$ and $\sigma \sim T/\om^2$ for $T \gg \om$. Next, within the $2-$pair excitation continuum, we show that the attenuation rate of a coherent mode of dispersion $\Omega_q$ crosses over from $\gamma_q \propto \Omega_q~(q/k_F)^2$, {\it e.g.} $\gamma_q \sim |q|^3$ for an acoustic mode, to $\gamma_q \propto T~(q/k_F)^2$, independent of $\Omega_q$, as temperature increases. Finally, we show that the $2-$pair excitation continuum yields subleading curvature corrections to the electron-electron scattering rate: $\tau^{-1} \propto V^2 T + V^4~T^3/\eps_F^2$, where $V$ is the dimensionless strength of the interaction.
\end{abstract}

\maketitle


\section{Introduction}
\label{introduction}

In 1933, Bloch~[\onlinecite{Bloch}] suggested that the low-energy excitations of an assembly of Fermi particles could be described in terms of "sound waves". Some years latter Tomonaga~[\onlinecite{Tomonaga}] provided a proof to Bloch's conjecture in the one-dimensional (1D) case. His demonstration was mainly~\cite{Note:FS} based on the linearization of the single-particle fermionic spectrum, $\xi_k^{\pm} = \pm v (k \mp k_F)$, where $\pm k_F$ are the Fermi points, $\pm$ correspond to the chiralities of the fermions: $+$ for right-movers near $+k_F$, $-$ for left-movers near $-k_F$ and $v$ the sound-wave velocity. On the basis of an equivalent model, of $(1+1)$ dimensional Dirac fermions, Luttinger [\onlinecite{Luttinger}] emphasized the non-Fermi liquid nature of 1D interacting fermions, {\it i.e.} the absence of fermionic quasi-particles. The relevancy of sound-waves of velocity $v$, as the correct low-energy (bosonic) quasi-particles (actually: particle-hole pairs), is reflected in the expression for the polarization operator, which reads:
\bea
\Pi_\pm^0(i\om,q) =  \pm {1 \ov 2\pi}{q \ov i\om \mp vq},~
\Pi^0(i\om,q) = {1 \ov \pi v} { (vq)^2 \ov (i\om)^2 - (vq)^2},
\nonum
\eea
where $\Pi^0(i\om,q) = \Pi_+^0(i\om,q) + \Pi_-^0(i\om,q)$.
The imaginary part of the retarded polarization operator defines the dynamical structure factor (DSF):
\be
S(\om, q) = - \Im \Pi^R(\om,q) = - \Im \Pi_+^R(\om,q) - \Im \Pi_-^R(\om,q).
\label{DSFdef}
\ee
For non-interacting fermions, in the linear spectrum approximation, Eq.~(\ref{DSFdef}) reads:
\be
S^{0}(\om, q) = {1 \ov v}~ (vq)^2 ~\delta[\om^2 - (vq)^2],
\label{DSFB}
\ee
which is a delta function centered around the spectrum $\om = \pm vq$ of the bosons. This result implies that these bosons have an infinite life-time and are free. Tomonaga has further shown that this result is valid in any order in the interaction among the fermions. Diagrammatically, this statement follows from the loop-cancellation theorem, which states that the fermionic loop with two external lines is the only non-zero one. That is, the sum of all loops with $3$ or more external lines is zero, {\it cf.} Ward identities due to Dzyaloshinskii and Larkin~[\onlinecite{DL}] which imply that the RPA is exact. Interactions therefore do not affect the coherent states to all orders which is the essence of the (non-perturbative in interactions) bosonization technique~[\onlinecite{Bosonization}]. The Tomonaga-Luttinger model together with the bosonization technique are the standard model and technique used to tackle a wide range of problems related to 1D interacting fermions, see the recent monographes~[\onlinecite{MonographeTG,MonographeAOG}].

Many solid-state problems require going beyond the Tomonaga approximation which assumes that fermions are of the Dirac-type with a "light" velocity, $v$. This amounts to take into account the fundamental feature of electrons in solids: band curvature, which should manifest at high enough energies (temperature, frequency, etc...). Within a fermionic approach this amounts to consider bare fermionic Green's functions of the form:
\be
G_{\pm}^0(i\eps,k) = {1 \ov i\eps - \xi_k^{\pm}}, \quad \xi_k^{\pm} = \pm v k + {k^2 \ov 2m},
\label{bare_green}
\ee
where the single particle fermionic spectrum has been expanded up to second order near the Fermi points $\pm k_F$; here and below: $k \equiv k \mp k_F$.
The academic problem of non-linear bosonization probably dates back to the times of Refs.~[\onlinecite{Tomonaga,DL,Bosonization}] and, since then, has attracted increasing attention, especially during the last decade or so, both in the field of the solid-state and in the more mathematically inclined literature, {\it e.g.} see Refs.~[\onlinecite{BMN,Kopietz,Samokhin,GlazmanDrag,Rozhkov,Wiegmann,PK,Glazman2,Affleck}]. This increase of focus on curvature effects is a witness of both their relevancy to contemporary physical applications as well as their complicated technical handling which delayed their quantitative analysis. The main physical motivations are related to the attenuation of 1D plasmons [\onlinecite{Samokhin,PK,Glazman2,Affleck}], small-momentum drag resistivity between quantum wires [\onlinecite{GlazmanDrag}] and shocks in 1D quantum waves associated with the gradient catastrophe of the corresponding non-linear partial differential equations as proposed in [\onlinecite{Wiegmann}]. Such phenomena do not manifest in the Tomonaga approximation. In particular, quantitative approaches to plasmon attenuation and small-momentum drag resistivity require a precise knowledge of the density-density correlation function and its dissipative part: the DSF, beyond the TL approximation; a major focus of this contribution. In this respect, the technical challenge is in handling the combined effects of curvature {\it and} interactions for this standard object. As has been shown above interactions alone can be taken into account straightforwardly by the bosonization technique in the Tomonaga-Luttinger model. The opposite limit of free-fermions is also worth considering because it allows a straightforward handling of band curvature and reveals some of its non-trivial features. The free-fermion polarization operator (at $T=0$) with curvature reads:
\begin{figure}
\includegraphics[width=5.5cm,height=3.5cm]{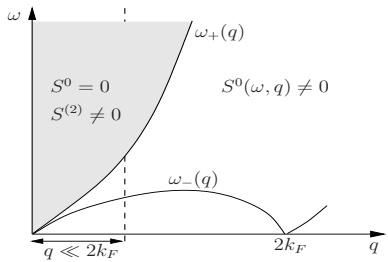}
\caption{ \label{Spectrum} Schematic view on the spectrum of excitations of 1D fermions. The free-fermion, or single-pair excitation continuum $S^0$, lies between the $\om_-$ and $\om_+$ branches. Because of strong energy-momentum constraints in 1D the dynamical structure factor is zero below the $\om_-$ branch (at finite temperature an exponentially small tail appears). We will focus on the multi-pair excitation continuum, which spreads above the $\om_+$ branch, at the $2-$pair level, $S^{(2)}$.}
\end{figure}

\begin{subequations}
\label{0MDSF}
\bea
&&\Pi^0(i\om,q)= { m \ov 2 \pi q}
\ln{ \left[{(i \om)^2 - (\om_-)^2 \ov (i\om)^2 - (\om_+)^2} \right]},
\label{PolarNL} \\
&& S^0(\om,q) = { m \ov 2 |q|} \left[ \theta [{mv \ov q}(\om - \om_-)] - \theta [{mv \ov q}(\om - \om_+)] \right]
\nonum \\
&&-~ \{ \om \rightarrow -\om\},
\label{DSFNL}
\eea
\end{subequations}
where the $2-$parametric family appears:
\be
\om_\pm (q) = vq \pm {q^2 \ov 2m}.
\label{2parametric}
\ee
Eq.~(\ref{DSFNL}) shows explicitly that the DSF satisfies~[\onlinecite{AGD}]: $S(\om,q) = - S(-\om,q),~~S(\om,q) =   S(\om,-q)$, together with the f-sum rule:
\be
\int_0^\infty {{d\om}} ~\om~ S(\om,q) = {v}~{q^2 \ov 2 \pi},
\label{FSR}
\ee
where the right-hand side in $m-$independent.

Curvature of the spectrum resulting from a finite mass, $m$, broadens the delta-function singularities in the range of frequencies: $\om_- < \om <  \om_+$ around $vq$ and $-\om_+ < \om < -\om_-$ around $-vq$. This yields a non-trivial spectrum of excitation, {\it cf.} Fig.~\ref{Spectrum}. The DSF therefore consists of two boxes centered around: $\om = \pm vq$, each box having a width $\delta \om = |\om_+ - \om_-| = q^2/m$ and a hight $\pm m/2|q|$ ($+$ sign for right-movers and $-$ sign for left-movers). Notice that at non-zero temperatures the Heaviside functions $\theta(x)$, in Eq.~(\ref{DSFNL}), become the Fermi occupation functions $n_F(x)$. At high temperatures, $T \gg v|q|$, the latter become the Boltzmann distribution functions which decay exponentially for $\om - v|q| \gg |q| T / k_F$ (we have used the fact that $k_F = m v$) providing some tails to the DSF. By definition, the $T=0$ (non-analytic) boxes and the $T\not=0$ (smooth) tails correspond to the incoherent single-pair excitation continuum whereby a coherent mode ({\it e.g.} the plasmon at $\om = vq$) decays by emitting a single particle-hole pair, see [\onlinecite{Nozieres}] for a review. The total width of the single-pair excitation continuum: $\delta \om = |\om_+ - \om_-| = q^2/m~+~|q| T / k_F$, corresponds to the decay rate of the plasmons in the free-fermion case. When $m \rightarrow \infty$ this decay rate goes to zero and the plasmons are again free in accordance with the Tomonaga-Luttinger result.

\begin{figure}
\includegraphics[width=7cm,height=3.5cm]{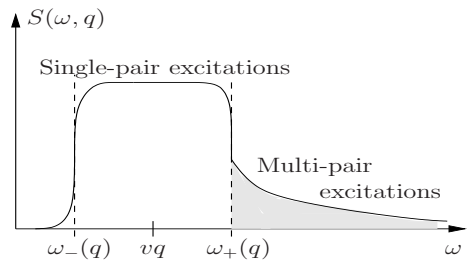}
\caption{ \label{DSF} Schematic view on the dynamical structure factor as a function of frequency $\om$ for a given momentum q and at finite temperatures, {\it e.g.} the cut at $q \ll 2k_F$ of Fig.~\ref{Spectrum}. The free-fermion or single-pair excitation continuum lies between the $\om_-$ and $\om_+$ branches. The $2-$ and higher-pair excitation continuum due to interactions lie beyond the $\om_+$ branch. We will focus on the $2-$pair excitation continuum, $\om \gg \om_+$. Effects of interactions below $\om_+$ are not represented.}
\end{figure}

More non-trivial effects on the shape of the DSF are expected from interactions among fermions, cf. Fig.~\ref{DSF}. At the text-book level, it is known that interactions yield an additional multi-pair excitation tail to the DSF, see [\onlinecite{Nozieres}]. In 1D such a continuum can appear only above the $|\om_+|-$branch due to strong energy-momentum constraints, see [\onlinecite{MonographeTG}], {\it i.e.} there is no dissipation below the $|\om_-|-$branch. In this frame, the authors of Ref.~[\onlinecite{GlazmanDrag}] have given an expression for the zero-temperature 2-pair excitation tail with the help of a fermion amplitude technique and related it to the small-momentum drag resistivity between quantum wires. More recently, the authors of Ref.~[\onlinecite{Affleck}] have derived the zero-temperature $2-$pair excitation tail and generalized it to all interaction strengths (full multi-pair excitation continuum) with the help of a spin-$1/2$ Heisenberg chain model (equivalent to spinless fermions, see [\onlinecite{MonographeTG,MonographeAOG}]) and its related bosonic representation. In this contribution, we also consider the $2-$pair excitation continuum and generalize it to non-zero temperatures by using diagrammatics which we think brings another different and interesting approach to the problem. We further apply our results to the computation of the optical conductivity, the damping rate of coherent excitations and to the electron-electron scattering rate. In each case, we attempt to firmly ground the relevancy of our methodology and results as well as compare them with the ones known from the literature on the subject.

The paper is structured as follows. In Sec.~\ref{diagrammatics_results} we justify the diagrammatic approach to the problem of the tails and deliver the results obtained with the help of this technique. In Sec.~\ref{applications} we apply such results to the calculation of the optical conductivity, the Landau damping rate of a coherent excitation and the electron-electron scattering rate. In Sec.~\ref{diagrammatics_calculations} we provide the reader with a demonstration of the results of Sec.~\ref{diagrammatics_results}. The conclusion and outlook are given in Sec.~\ref{conclusion}.

\section{Diagrammatic approach}
\label{diagrammatics_results}

\subsection{Motivation}

The major difficulty arising from diagrammatics is that the interplay between curvature and interactions leads to the breakdown of the loop cancellation theorem, a substantial technical complication as diagrams beyond the RPA are generated. Even though difficult to re-sum, we think this approach provides a rather systematic and controlled way to compute correlation functions. The basic reason behind this statement is the fact that the fermionic approach takes fully into account of the 2-parametric family $\{\om_-, \om_+\}$, cf. Eq.~(\ref{2parametric}). On the other hand, (linear) bosonization maps this family to the single-parametric one, i.e. the bosons of Bloch. In order to emulate a 2-parametric family from a single one, within bosonization, non-linear terms have to be added to the usual bosonized action, {\it e.g.} $(\partial_x \vp)^3$ in [\onlinecite{Samokhin}] for fermion curvature and $(\partial_t \vp)^4$ in [\onlinecite{BD}] for phonon curvature. This is what makes the bosonization non-linear. It turns out that these terms are highly singular even at the second order of perturbation theory in curvature (singularity of the $\delta(0)-$type, cf. [\onlinecite{Samokhin,BD}]). This singularity signals that Bloch's "sound" is highly non-trivial as it corresponds to the merging of {\it two} branches. The price to pay is the need to rely on a self-consistent approximation to cure the singular perturbation theory, cf. [\onlinecite{Samokhin,BD}]. Once this is done interactions may be taken into account in all orders which allows to explore some non-perturbative features of the interacting fermions. For example, it has been shown in Ref.~[\onlinecite{Samokhin}] that the $T=0$ decay rate of the plasmons reads: $\gamma_q \propto q^2/m^*$, where $m^*$ is renormalized by the interactions among the fermions and goes to the band mass $m$ when interactions are switched off (in agreement with the free-fermion case). On the other hand, sharp features of the DSF, {\it e.g.} the (smooth) multi-particle tail which transits to a (non-analytic) box-like shaped continuum in the $T=0$ free-fermion limit, are not clearly revealed by the self-consistent approximation, as far as we could learn from Ref.~[\onlinecite{Samokhin}]. These issues seem to have been resolved in Ref.~[\onlinecite{Affleck}]. As for our concerns, the advantage of the diagrammatic technique is precisely in that it avoids the self-consistent approximation. Moreover, it is particularly well suited for the computation of the 2-pair excitation tail (the decay of high-frequency coherent modes into 2 particle-hole pairs) as the latter is fully determined by a second-order perturbation theory in the interaction among the fermions. Finally, if re-summed correctly, it would even allow to explore non-perturbative features of the fermions and thus be fully complementary to the self-consistent non-linear bosonization of Ref.~[\onlinecite{Samokhin}].

\subsection{Model}

\begin{figure}
\includegraphics[width=8cm,height=6cm]{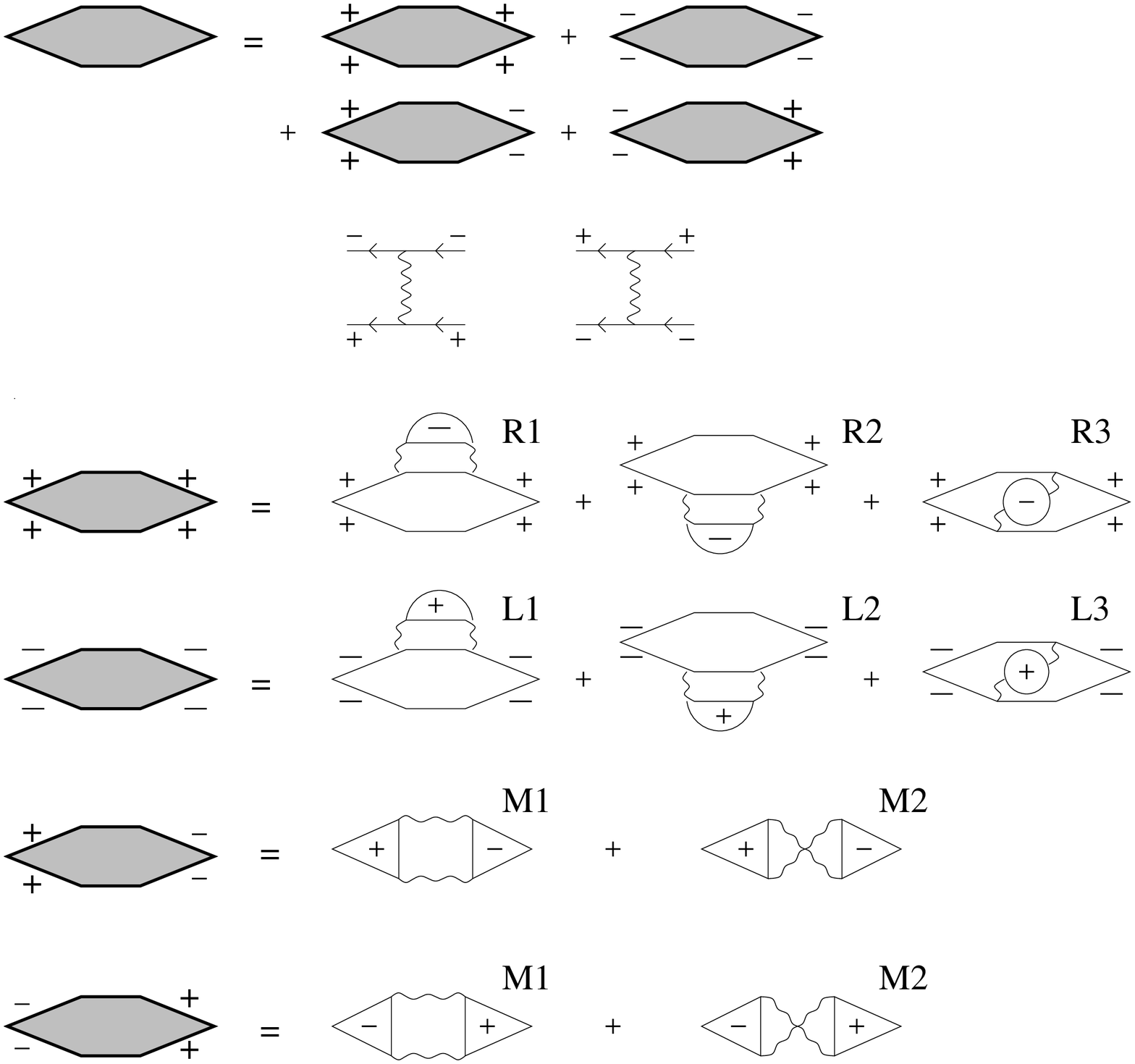}
\caption{ \label{Diagrams2nd} The 2nd order contributions to the polarization operator of spinless fermions interacting with the $g_2$ process, {\it i.e.} $+$ fermions scattering on $-$fermions and vice-versa. They are divided into three classes: two "chiral" classes (R and L) and one "mixed" class (M). All these diagrams correspond to the decay of a density fluctuation, $\{\om,q\}$, into two particle-hole pair excitations. They describe the microscopic scattering mechanisms taking place within the $2-$pair excitation continuum. }
\end{figure}

Given this important justification regarding our working tool, the simplest model which captures the interplay between curvature and interactions is the one of spinless fermions interacting with the $g_2$ process, {\it i.e.} right-moving ($+$) fermions scattering on left-moving ($-$) fermions and vice-versa. This model neglects the $g_4-$ process~\cite{g4_diagrams} as well as the $g_3-$ or Umklapp process, so we are away from half-filling, see Ref.~[\onlinecite{Affleck}] for a more complete theory. Within this model, curvature is fully taken into account while interactions are treated perturbatively. In the second order 10 diagrams arise, {\it cf.} Fig.~\ref{Diagrams2nd}. These diagrams are well-known and one can check, from Fig.~\ref{Diagrams2nd}, that they indeed correspond to the decay of a density fluctuation, $\{\om,q\}$, into two particle-hole pair excitations. They may be divided into three classes: two "chiral" classes (R and L) and one "mixed" class (M); this terminology will be justified a bit latter. Diagrams R1, R2, L1 and L2 correspond to the re-normalization of a particle line. Diagrams R3 and L3 to a vertex correction. Diagrams M1 and M2 (which appear with a factor of two) are known in the theory of superconductivity as the Aslamasov-Larkin diagrams and in mesoscopics as the Altshuler-Shklovskii diagrams. To the knowledge of the author, these diagrams have not been computed for 1D fermions with mass. In 3D, the calculation was done by DuBois and Kivelson~[\onlinecite{DuBois}] and in 2D by Mishchenko et al. following the work of Reizer and Vinokur~[\onlinecite{Reizer1}]. As previously mentioned, in 1D, the only fermionic approach we are aware of is the one of Ref.~[\onlinecite{GlazmanDrag}] concerning the tails of the DSF. However, the expression of the tails given in Ref.~[\onlinecite{GlazmanDrag}] is based on summing fermion amplitudes. This procedure generates R and L diagrams but neglects M diagrams. As we will show in the following, M diagrams are crucial to derive the tails, {\it i.e.} they bring some cancellations with L and R diagrams which correctly enforce the satisfaction of the f-sum rule, Eq.~(\ref{FSR}). Given these methodological considerations, our final $T=0$ result agrees with the one of [\onlinecite{GlazmanDrag}] and of the more recent~[\onlinecite{Affleck}]. We now outline our main results.

\subsection{Results}

1) We consider first the case $m \rightarrow \infty$. From the loop cancellation theorem, the sum of all diagrams should be zero. We find that the combinations:  R1+R2+R3, L1+L2+L3 and M1+M2, vanish in this limit. This implies that all R diagrams contribute equally, as well as all L diagrams and both M diagrams. Instead of working with individual diagrams we therefore work with the above combinations.

2) We consider then right- and left-movers with curved spectrum. Working with the combinations of diagrams defined in 1), which vanish in the case of a linear spectrum, we implement a systematic expansion in $1/m$. We prove that diagrams R1+R2+R3 have an expansion in $1/m_+^2$, diagrams L1+L2+L3 have an expansion in $1/m_-^2$ and diagrams M1+M2 have an expansion in $1/m_+ m_-$ ($m_+$ and $m_-$ being the masses of right- and left-movers, respectively). This justifies the denomination of R and L diagrams as chiral diagrams and M diagrams as mixed diagrams (they mix the chiralities of the fermions).

3) We focus on the lowest order in the above $1/m$ expansion. This amounts to focus on the tails of the DSF, $|\om| \gg vq$.
The arguments of 2) show that the tails provided by R (resp. L) diagrams correspond to the scattering of a curved right- (resp. left-) mover on a linear left- (resp. right-) mover. On the other hand the tails of the mixed diagrams require the scattering of two curved fermions with different chiralities. This can be easily seen from our general results which read:
\begin{subequations}
\label{Results}
\bea
&&\Im \Pi_R^{(2)R}(\om,q) = - {|V[{\om-vq \ov 2}]|^2 \ov 256 \pi v^5}~{q^2 \ov m_+^2}~
{\om + vq \ov \om - vq}~{\mathcal{F}(T;\om,q)} \nonum \\
&&- \left \{ \om \rightarrow -\om, \quad q \rightarrow -q \right \},
\label{R} \\
&&\Im \Pi_L^{(2)R}(\om,q) = - {|V[{\om+vq \ov 2}|^2 \ov 256 \pi v^5}~{q^2 \ov m_-^2}~{\om - vq \ov \om + vq}~{\mathcal{F}(T;\om,q)} \nonum \\
&&- \left \{ \om \rightarrow -\om, \quad q \rightarrow -q \right \},
\label{L} \\
&&\Im \Pi_{M}^{(2)R}(\om,q) = + {V[{\om+vq \ov 2}] V[{\om-vq \ov 2}] \ov 256 \pi v^5}~{2 q^2 \ov  m_+ m_-}~{\mathcal{F}(T;\om,q)}
\nonum \\
&&- \left \{ \om \rightarrow -\om, \quad q \rightarrow -q \right \},
\label{M}
\eea
\end{subequations}

\ni where $\Im \Pi_\eta^{(2)R}(\om,q)$ is the sum of all diagrams in the given class $\eta=R,L,M$ and the temperature factor reads:
\be
{\mathcal{F}}(T;\om,q) = {1 \ov 1 - e^{-{\om - vq \ov 2T}}} - {1 \ov 1 - e^{{\om + vq \ov 2T}}}.
\label{TFactor}
\ee

4) We first comment on the effect of the long-range potential $V[q]$.
As these expressions are valid for the tails of the correlation function, $|\om| \gg vq$, the interaction potential factorizes and brings a contribution: $|V[\om/2]|^2$. The three groups of diagrams (R, L and M) therefore cannot be discriminated by the inclusion of a long-range potential as far as the tails are concerned. Notice, however, that the long-range static Coulomb interaction acquires a dynamical nature depending on the external frequency, $\om$. This dependence provides further $\om-$tails to the dynamical structure factor. Actually, for a realistic three-dimensional Coulomb interaction we have: $V(q) = -e^2 \ln(|q|a)$, where $a$ is a short distance cut-off. Therefore, the long-range 3D Coulomb potential provides an additional frequency dependence: $|V[\om/2]|^2 = e^4 \ln^2(|\om|a/2v)$, which is marginal.

5) Next we comment on the effect of temperature. Notice that the temperature factor of Eq.~(\ref{TFactor}) is common to all diagrams and therefore disables any discrimination between them. It has the following limiting expressions:
\begin{subequations}
\bea
&&{\mathcal{F}}_0 = \Theta[\om - vq] - \Theta[-\om -vq], \quad |\om \pm vq| \gg T,~~~
\label{TFactor0} \\
&&{\mathcal{F}}_\infty = {4T \om \ov \om^2 -(vq)^2}, \quad |\om \pm vq| \ll T.
\label{TFactorT}
\eea
\end{subequations}
We see clearly from the zero-temperature result of Eq.~(\ref{TFactor0}) that the tails are restricted to the regions $|\om| \gg vq$, {\it i.e.} beyond the $|\om_+|-$line in agreement with Refs.~[\onlinecite{GlazmanDrag}] and~[\onlinecite{Affleck}]. For $T \gg |\om|$, the tails acquire a linear $T-$dependence, {\it cf.} Eq.~(\ref{TFactorT}). At this point we should keep in mind the existence of a tail due to the single-pair excitation continuum. As proved in the Sec.~\ref{introduction}, this tail decays exponentially for: $\om -v|q| \gg |q|T/k_F$. This implies that the 2-pair excitation continuum dominates in the range of frequencies: $|q|T/k_F \ll \om -v|q| \ll T$ at finite temperatures. This contribution is substantial for $|q| \ll k_F$.

6) We focus on the specific case of zero-temperature. In the following we assume that $m_+ = m_- = m$. We also consider a point-like interaction, $V[q]=V_0$, even though all results may be generalized to the case of a long-range interaction by changing: $V_0 \rightarrow V[\om/2]$. We find that, as far as the tails of the DSF are concerned, all groups of diagrams contribute equally to the correlation function:
\be
S^{(2)}(\om,q) = - \Im \Pi_{R}^{(2)R}(\om,q) - \Im \Pi_{L}^{(2)R}(\om,q) - \Im \Pi_{M}^{(2)R}(\om,q),
\ee
which reads:
\bea
S^{(2)}(\om,q) =&& {V_0^2 \ov 64 \pi v^3}~\left({q^2 \ov m}\right)^2~{1 \ov \om^2 - (vq)^2}~{\mathcal{F}}_0
\nonum \\
&& -~\left \{ \om \rightarrow -\om, \quad q \rightarrow -q \right \}, \quad T \ll |\om|,~~~~~~
\label{DSF0T}
\eea
where ${\mathcal{F}}_0$ is given by Eq.~(\ref{TFactor0}). Contrary to Eqs.~(\ref{R}) and (\ref{L}) as well as their sum, Eq.~(\ref{DSF0T}) satisfies Eq.~(\ref{FSR}) and agrees with the result of Refs.~[\onlinecite{GlazmanDrag}] and~[\onlinecite{Affleck}].

7) We focus now on finite temperatures, {\it i.e.} for $T \gg |\om|$ (we still have $|\om| \gg vq$). Proceeding along the lines of the zero-temperature case yields:
\bea
S^{(2)}(\om,q) =&& {V_0^2 \ov 16 \pi v^3}~\left({q^2 \ov m}\right)^2~{\om~T \ov (\om^2 - (vq)^2)^2}
\nonum \\
&&-~\left \{ \om \rightarrow -\om, \quad q \rightarrow -q \right \}, \quad T \gg |\om|,~~~~
\label{DSFT}
\eea
which satisfies Eq.~(\ref{FSR}). Temperature therefore gives rise to frequency-tails proportional to $T / \om^3$ for a point-like interaction. Such 2-pair excitation tails dominate the single-pair tail in the range of frequencies: $|q|T/k_F \ll \om -v|q| \ll T$, as has been shown above.

8) A general equation which interpolates between Eq.~(\ref{DSF0T}) and Eq.~(\ref{DSFT}) reads:
\bea
S^{(2)}(\om,q) =&& {V_0^2 \ov 64 \pi v^3}~\left({q^2 \ov m}\right)^2~{1 \ov \om^2 - (vq)^2}~{\mathcal{F}}(T;\om,q)
\nonum \\
&& -~\left \{ \om \rightarrow -\om, \quad q \rightarrow -q \right \},
\label{DSFtails}
\eea
where the temperature factor is given by Eq.~(\ref{TFactor}).

9) Finally, we comment on the approach to the single-pair continuum: $\om \rightarrow \pm vq$. As can be seen from Eqs.~(\ref{Results}), in the limit $\om \rightarrow + vq$, Eq.~(\ref{R}) is singular whereas Eq.~(\ref{L}) vanishes. On the other hand, in the limit $\om \rightarrow - vq$, Eq.~(\ref{L}) is singular whereas Eq.~(\ref{R}) vanishes. These singularities are re-enforced by the divergency of the long-range Coulomb interaction within the same limits. In all cases, Eq.~(\ref{M}) is either constant or diverges less strongly than the corresponding relevant chiral part in the presence of a long-range potential and/or finite temperatures. This implies that, even though M diagrams are crucial for the f-sum rule to be satisfied, the chiral diagrams become more relevant (divergent) than them near the singular line.

\section{Applications}
\label{applications}

\subsection{Optical conductivity}

As a first application we determine the interaction correction to the tails of the conductivity (the optical conductivity) beyond the Tomonaga approximation. The latter is defined as:
\be
\Re \sigma(\om,q) = e^2~{\om \ov q^2}~S(\om,q).
\label{sigmadef}
\ee
In the absence of curvature Eq.~(\ref{DSFB}) shows that the conductivity consists of Drude peaks at $\om =\pm vq$. In the following we focus only on the peak at: $\om = + vq$. Including curvature from Eq.~(\ref{DSFNL}) yields:
\bea
&&\Re \sigma^0(\om,q) =
\nonum \\
&&{e^2 v \ov 2}~{\om \tau_q \ov vq}~\left[ n_F[-{mv \ov q}(\om - \om_-)] - n_F[-{mv \ov q}(\om - \om_+)] \right],
\nonum
\eea
where $\tau_q = m/q^2$ is the decay-time of the Bloch-Tomonaga-Luttinger bosons within the free-fermion continuum. At low temperatures, the Fermi functions reduce to step functions and the above conductivity is non-zero only within the continuum: $\om_- < \om < \om_+$ (idem for negative frequencies). At finite temperatures, $T \gg vq$, the Fermi functions become the Boltzmann functions which yield additional tails to the conductivity, $\propto \exp[-mv (\om -v|q|) / |q|T]$. These tails manifest for: $\om -v|q| \ll |q|T/mv$ and are exponentially suppressed beyond.

Including interactions, our previous results yield the interaction-correction to the high-frequency ($|\om| \gg vq$) small-momentum ($q \ll 2k_F$) conductivity:
\begin{subequations}
\label{sigma}
\bea
\Re \sigma^{(2)}(\om,q) =&& {e^2 V_0^2 \ov 64 \pi v^3}~{q^2 \ov m^2}~{\om \ov \om^2 - (vq)^2}~{\mathcal{F}}(T;\om,q).
\label{sigmatails} \\
\Re \sigma^{(2)}(\om,q) =&& {e^2 V_0^2 \ov 64 \pi v^3}~{q^2 \ov m^2}~{1 \ov \om},~~~T \ll \om,
\label{sigmatails_0T} \\
\Re \sigma^{(2)}(\om,q) =&& {e^2 V_0^2 \ov 16 \pi v^3}~{q^2 \ov m^2}~{T \ov \om^2},~~~T \gg \om.
\label{sigmatails_T}
\eea
\end{subequations}
The $\om-$tails are $\propto 1/\om$ for $\om \gg T$ and $\propto T/\om^2$ for $\om \ll T$. At high temperatures there is therefore a fairly large regime of frequencies: $qT/k_F \ll \om -v|q| \ll T$, where curvature corrections provide the dominant contribution to the conductivity.

Such an optical conductivity can be accessed experimentally in semi-conducting quantum wires and organic charge-density waves, {\it e.g.} Bechgaard salts. The latter have very rich properties due to, {\it e.g.} the presence of an inter-chain coupling: $t_\perp$ and commensurability effects. At high enough temperatures (above the dimensional crossover: $T \gg t_\bot$) the chains decouple and 1D theories may apply. Based on the proximity to a Mott insulator (due to Umklapp processes) the authors of Refs.~[{\onlinecite{GiamarchiOrganics}] provided a theory for the optical conductivity of the Bechgaard salts related to recent experiments in the field. In this respect, we would simply like to point out that, if Umklapp processes are negligible, band-curvature allows forward-scattering processes to be considered as an alternative non-trivial source of optical conductivity, {\it cf.} Eq.~(\ref{sigma}).

\subsection{Attenuation of coherent excitations}

Next we focus on the damping rate, $\gamma_q$ of coherent modes. This will push to its very limits the perturbative approach that we have implemented. However, we find the obtained results interesting enough to be reported and we will comment on further developments in the Conclusion, Sec.~\ref{conclusion}. Experimentally, this attenuation may be accessed via energy-loss experiments, {\it e.g.} see [\onlinecite{experiments}].

\begin{figure}
\includegraphics[width=7cm,height=2.5cm]{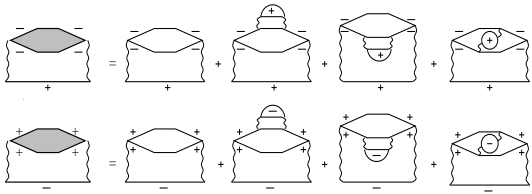}
\caption{ \label{Diagrams2ndGF} Self-energy diagrams for spinless fermions interacting with the $g_2$ process. The second-order diagrams are usual Luttinger-liquid diagrams. The next three diagrams are of fourth-order in interaction among the fermions. They are built from the polarization parts of the previous section and yield the lowest non-trivial curvature corrections: second order in $1/m$ to, {\it e.g.} the single-particle scattering rate due to electron-electron interactions.}
\end{figure}

Within a naive RPA approximation the plasmon dispersion relation and the Landau damping rate are defined with the help of the following equations:
\bea
\label{coherent_mode}
&& 1 = V(q)~\Re \Pi^R(q,\omega_q),
\label{dispersion} \\
&& \gamma_q = \Im \Pi^R(q,\omega_q)~[ \partial_\om \Re \Pi^R(q,\omega_q) ]^{-1},
\label{attenuation}
\eea

\ni where: $\Pi = \Pi^0 + \Pi_R + \Pi_L + \Pi_M$, $V(q) = -e^2 \ln(|q|a)$ and we have assumed that the damping is small, $\gamma_q \ll \omega_q$. With the help of the non-interacting result of Eq.~(\ref{0MDSF}), the 1D plasmon branch is given by~\cite{Note:plasmon_dispersion}: $\omega_q = v |q|$, in the lowest order in $1/m$ and falls precisely in the singular limit of our perturbation theory, as we already know from previous sections. A non-trivial re-summation procedure is necessary in order to rigorously take into account of the effect of interactions on this damping rate (recall that for free fermions the damping rate is simply given by: $\gamma_q = q^2/m$). This is a basic difference with higher dimensions where one can straightforwardly account for the damping within perturbation theory, cf. [\onlinecite{DuBois,Reizer1}].

This problem was the focus of the recent literature, see Refs.~[\onlinecite{PK,Glazman2,Affleck}]. We will focus on a slightly different issue by considering a coherent mode of frequency $\Omega_q$ lying deep within the $2-$pair excitation spectrum, $\Omega_q \gg \om_+$. In the limit where $\Omega_q \rightarrow \om_q$ this becomes a proper mode of the system and satisfies Eq.~(\ref{dispersion}). Because of the $2-$pair continuum some energy of this mode is absorbed by the system and leads to its attenuation according to Eq.~(\ref{attenuation}) with $\om_q = \Omega_q$, $\Im \Pi^R(\om,q) = -S^{(2)}(\om,q)$ of Eq.~(\ref{DSFtails}) and $\Re \Pi^R(\om,q) = \Re \Pi^{R0}$ of Eq.~(\ref{PolarNL}). The attenuation of this high-frequency coherent mode due to the $2-$pair excitation continuum is then given by:
\bea
\gamma_q = \left({V(q) \ov v}\right)^{2}~\left({q \ov m v}\right)^2~\Omega_q~{\mathcal{F}}(T,\Omega_q),
\label{damping}
\eea
up to a numerical factor and in the lowest order in $1/m$. At zero temperature, Eq.~(\ref{damping}) yields:
\bea
\gamma_q = \left( {V(q) \ov v} \right)^{2}~\left({q \ov m v}\right)^2~\Omega_q,~~T \ll \Omega_q.
\label{damping_0T}
\eea
Including temperature, Eq.~(\ref{damping}) yields:
\bea
\gamma_q = \left( {V(q) \ov v} \right)^{2}~\left({q \ov m v}\right)^2~T, ~~T \gg \Omega_q,
\label{damping_T}
\eea
which does not depend on $\Omega_q$ and dominates the damping due to the finite $T$ single-pair excitation continuum for: $|q| T / k_F \ll \Omega_q \ll T$.
A simple formula interpolating between Eqs.~(\ref{damping_0T}) and (\ref{damping_T}) reads:
\bea
\gamma_q = \left( {V(q) \ov v} \right)^{2}~\left({q \ov m v}\right)^2~{\mathrm{max}}\{~T,~\Omega_q~\}.
\label{damping_interpolation}
\eea
Notice that if the mode is acoustic: $\Omega_q \sim |q|$, then the attenuation rate crosses over from $\gamma_q \sim |q|^3$ at small temperatures to $\gamma_q \sim q^2$ at high temperatures.

\subsection{Electron-electron scattering rate}

Finally, we determine curvature corrections to the electron-electron scattering rate. It corresponds to the imaginary part of the self-energy diagrams represented on Fig.~\ref{Diagrams2ndGF} on the Fermi shell, $\tau_\pm^{-1}(\eps) \equiv \Im \Sigma_{\pm \mp}(\eps,k=0)$. Notice that, because it is a single-particle property, the scattering rate does not depend on the mixed component of the polarization part. The second order self-energy parts are well-defined in the massless case and yield: $\tau_{\pm}^{(2)-1} = - (V_0/v)^2~{\mathrm{max}} \{|\eps|,2T\}$, for a point-like interaction. The linear dependence on energy is known, {\it e.g.} see Ref.~[\onlinecite{Maslov}], and is another signature of the absence of fermionic quasi-particles at low energies~\cite{Note:FL}. From Eqs.~(\ref{R}) and~(\ref{L}) we may now access {\it corrections} to this scattering rate which go beyond the Tomonaga approximation: $\tau^{-1} = \tau^{(2)-1} + \tau^{(4)-1} +...$. The lowest-order corrections are of the fourth-order in interaction and second-order in mass:
\bea
&&\tau^{(4)-1} = - {V_0^4 \ov v^4}~{T^3 \ov \eps_F^2},
\label{scatt4}
\eea
up to a numerical factor. At low-energies, Eq.~(\ref{scatt4}) is small in comparison with the Luttinger-liquid contribution: $\tau^{(2)-1} \sim T$. This result is correct up to the fact that we only know the tails of the DSF. Notice that this correction does not follow the usual, Fermi-liquid-like, scaling: $T^2 / \eps_F$. However, the cubic dependence on energy seems natural because the mass-dependence of Eq.~(\ref{scatt4}) -- recall that: $\eps_F = m v^2$ -- originates from the polarization parts appearing in Fig.~\ref{Diagrams2ndGF}, {\it i.e.} the ones analyzed in detail in Sec.~\ref{diagrammatics_results}. From dimensional arguments we understand the cubic energy-dependence as the only possible scaling to compensate for the square of the Fermi energy appearing from the tails of the polarization parts.

\section{Diagrammatic calculations}
\label{diagrammatics_calculations}

In this section we provide the interested reader with some details regarding the diagrammatic calculations. All the calculations are based on the fact that the bare Green's function, $G^0$, is the one of Eq.~(\ref{bare_green}). Moreover, one should keep in mind the loop cancellation theorem of Dzyaloshinskii and Larkin which has also been defined in the introduction. This theorem is equivalent to Furry's theorem in quantum electro-dynamics (QED) so what we are doing in the language of QED is to calculate non-trivial radiative corrections which are unique to solid-state fermions.

The calculation is then divided into two parts: the first part concerns the "chiral" classes (R and L) and the second part concerns the "mixed" class (M).

\subsection{Chiral (R and L) diagrams with curvature}

A convenient starting point is to work with the sums of diagrams which combine to vanish in the massless limit, i.e. R1 + R2 + R3 (idem for L diagrams). Re-arranging the Green's functions in the expressions of such diagrams splits them into simpler expressions and generates a systematic expansion in $1/m$.

\begin{widetext}

The expressions for the 3 diagrams are the following:
\bea
&&\Pi_{R1}^{(2)}(i\om,q) = {1 \ov \beta^2} \sum_{\eps,\om'} \int {{dk dq'\ov (2\pi)^2}}~|V[q']|^2~\Pi_-^0(i\om',q')
~G_{+}^{0}(i\eps,k) \left[ G_{+}^{0}(i(\eps-\om),k-q) \right]^2 G_{+}^{0}(i(\eps-\om+\om'),k-q+q'),
\nonum \\
&&\Pi_{R2}^{(2)}(i\om,q) = {1 \ov \beta^2} \sum_{\eps,\om'} \int {{dk dq'\ov (2\pi)^2}}~|V[q']|^2~\Pi_-^0(i\om',q')
~\left[ G_{+}^{0}(i\eps,k) \right]^2 G_{+}^{0}(i(\eps-\om),k-q) G_{+}^{0}(i(\eps+\om'),k+q'),
\nonum \\
&&\Pi_{R3}^{(2)}(i\om,q) = {1 \ov \beta^2} \sum_{\eps,\om'} \int {{dk dq'\ov (2\pi)^2}}~|V[q']|^2~\Pi_-^0(i\om',q')
\nonum \\
&&G_{+}^{0}(i\eps,k) G_{+}^{0}(i(\eps-\om),k-q) G_{+}^{0}(i(\eps + \om'),k + q')G_{+}^{0}(i(\eps-\om+\om'),k-q+q'),
\nonum
\eea
where the double integrals over momentum show that these are $2-$loop contributions. The expression for R3 may be further split into three contributions because of the following identity:
\bea
G_{+}^{0}(i(\eps + \om'))G_{+}^{0}(i(\eps-\om+\om')) =&&
{1 \ov i\om -q}~[~G_{+}^{0}(i(\eps-\om+\om')) - G_{+}^{0}(i(\eps + \om'))
\nonum \\
&& + (\xi_{k+q'}^{(+)} - \xi_{k+q'-q}^{(+)} - q)G_{+}^{0}(i(\eps-\om+\om'))G_{+}^{0}(i(\eps + \om'))~]
\nonum
\eea
where the last term is a clear manifestation of the non-linear spectrum. This yields:
\bea
&&\Pi_{R3}^{(2)}(i\om,q) = {1 \ov i\om - q}~{1 \ov \beta^2} \sum_{\eps,\om'}
\int {{dk dq'\ov (2\pi)^2}}~|V[q']|^2~\Pi_-^0(i\om',q')~[
\nonum \\
&&G_{+}^{0}(i\eps) G_{+}^{0}(i(\eps-\om)) G_{+}^{0}(i(\eps-\om+\om'))
- G_{+}^{0}(i\eps) G_{+}^{0}(i(\eps-\om)) G_{+}^{0}(i(\eps + \om'))
\nonum \\
+&&(\xi_{k+q'}^{(+)} - \xi_{k+q'-q}^{(+)} - q)
G_{+}^{0}(i\eps) G_{+}^{0}(i(\eps-\om)) G_{+}^{0}(i(\eps-\om+\om')) G_{+}^{0}(i(\eps + \om'))~].
\nonum
\eea
We display this expression in a symmetric form (change of variables: $\eps \rightarrow \eps-\om$, $k \rightarrow k-q$ in the first term and $\eps +\om' \rightarrow \eps$, $k+q' \rightarrow k$ in the last with $\om' \rightarrow -\om'$, $q' \rightarrow -q'$):
\bea
&&\Pi_{R3}^{(2)}(i\om,q) = {1 \ov i\om - q}~{1 \ov \beta^2} \sum_{\eps,\om'}
\int {{dk dq'\ov (2\pi)^2}}~|V[q']|^2~\Pi_-^0(i\om',q')~[
\nonum \\
&&G_{+}^{0}(i\eps) G_{+}^{0}(i(\eps+\om)) G_{+}^{0}(i(\eps + \om'))
-G_{+}^{0}(i\eps) G_{+}^{0}(i(\eps-\om)) G_{+}^{0}(i(\eps + \om'))
\nonum \\
+&&{\xi_{k}^{(+)} - \xi_{k-q}^{(+)} - q \ov 2}
G_{+}^{0}(i\eps) G_{+}^{0}(i(\eps-\om)) G_{+}^{0}(i(\eps-\om+\om')) G_{+}^{0}(i(\eps + \om'))
\nonum \\
+&&{\xi_{k+q}^{(+)} - \xi_{k}^{(+)} - q \ov 2}
G_{+}^{0}(i\eps) G_{+}^{0}(i(\eps+\om)) G_{+}^{0}(i(\eps+\om+\om')) G_{+}^{0}(i(\eps + \om'))~].
\nonum
\eea
Notice that the change $\om' \rightarrow -\om'$, $q' \rightarrow -q'$ does not affect the interaction potential.
Also, the last term has been symmetrized.

After a change of variables in R1 ($\eps - \om \rightarrow \eps$, $k-q \rightarrow k$), the sum of R1 and R2 may be re-written as:
\bea
&&\Pi_{R1}^{(2)}(i\om,q) + \Pi_{R2}^{(2)}(i\om,q) = {1 \ov \beta^2} \sum_{\eps,\om'} \int {{dk dq'\ov (2\pi)^2}}~|V[q']|^2~\Pi_-^0(i\om',q')
\left [ G_{+}^{0}(i\eps) \right ]^2~G_{+}^{0}(i(\eps+\om')) \left \{ G_{+}^{0}(i(\eps-\om))+ G_{+}^{0}(i(\eps+\om)) \right \}.
\nonum
\eea
This expression shows that diagram R2 is equal to diagram R1 upon changing $\om \rightarrow -\om$ and $q \rightarrow -q$.
This can be re-written as:
\bea
&&\Pi_{R1}^{(2)}(i\om,q) + \Pi_{R2}^{(2)}(i\om,q) = {1 \ov \beta^2} \sum_{\eps,\om'} \int {{dk dq'\ov (2\pi)^2}}~|V[q']|^2~\Pi_-^0(i\om',q')~[
\nonum \\
&&2G_{+}^{0}(i\eps) G_{+}^{0}(i(\eps-\om)) G_{+}^{0}(i(\eps+\om)) G_{+}^{0}(i(\eps+\om'))
\nonum \\
&&+(2 \xi_k ^{(+)}- \xi_{k+q}^{(+)} - \xi_{k-q})
\left [ G_{+}^{0}(i\eps) \right ]^2 G_{+}^{0}(i(\eps-\om)) G_{+}^{0}(i(\eps+\om)) G_{+}^{0}(i(\eps+\om'))~].
\nonum
\eea
The sum of all three diagrams:
\be
\Pi_{R}^{(2)}(i\om,q) = \Pi_{R1}^{(2)}(i\om,q) + \Pi_{R2}^{(2)}(i\om,q) + \Pi_{R3}^{(2)R3}(i\om,q),
\ee
yields:
\bea
&& \Pi_{R}^{(2)}(i\om,q) = {1 \ov \beta^2} \sum_{\eps,\om'} \int {{dk dq'\ov (2\pi)^2}}~|V[q']|^2~\Pi_-^0(i\om',q')~[
\nonum \\
&&G_{+}^{0}(i\eps) G_{+}^{0}(i(\eps+\om)) G_{+}^{0}(i(\eps+\om'))
\left[ G_{+}^{0}(i(\eps-\om)) + {1 \ov i\om - q} \right]
\nonum \\
+&&G_{+}^{0}(i\eps) G_{+}^{0}(i(\eps-\om)) G_{+}^{0}(i(\eps+\om'))
\left[ G_{+}^{0}(i(\eps+\om)) - {1 \ov i\om - q} \right]
\nonum \\
+&&(2 \xi_k^{(+)} - \xi_{k+q}^{(+)} - \xi_{k-q}^{(+)})
\left [ G_{+}^{0}(i\eps) \right ]^2 G_{+}^{0}(i(\eps-\om)) G_{+}^{0}(i(\eps+\om)) G_{+}^{0}(i(\eps+\om'))
\nonum \\
+&&{\xi_{k}^{(+)} - \xi_{k-q}^{(+)} - q \ov 2(i\om - q)}
G_{+}^{0}(i\eps) G_{+}^{0}(i(\eps-\om)) G_{+}^{0}(i(\eps-\om+\om')) G_{+}^{0}(i(\eps + \om'))
\nonum \\
+&&{\xi_{k+q}^{(+)} - \xi_{k}^{(+)} - q \ov 2(i\om - q)}
G_{+}^{0}(i\eps) G_{+}^{0}(i(\eps+\om)) G_{+}^{0}(i(\eps+\om+\om')) G_{+}^{0}(i(\eps + \om'))~].
\nonum
\eea
This is further simplified as:
\bea
&&\Pi_{R}^{(2)}(i\om,q) = {1 \ov \beta^2} \sum_{\eps,\om'} \int {{dk dq'\ov (2\pi)^2}}~|V[q']|^2~\Pi_-^0(i\om',q')~[
\nonum \\
&&{\xi_{k+q}^{(+)}-\xi_{k-q}^{(+)}-2q \ov i\om - q}
~G_{+}^{0}(i\eps) G_{+}^{0}(i(\eps-\om)) G_{+}^{0}(i(\eps+\om)) G_{+}^{0}(i(\eps+\om'))
\nonum \\
+&&(2 \xi_k^{(+)}- \xi_{k+q}^{(+)} - \xi_{k-q}^{(+)})
~\left [ G_{+}^{0}(i\eps) \right ]^2 G_{+}^{0}(i(\eps-\om)) G_{+}^{0}(i(\eps+\om)) G_{+}^{0}(i(\eps+\om'))
\nonum \\
+&&{\xi_{k}^{(+)} - \xi_{k-q}^{(+)} - q \ov 2(i\om - q)}
~G_{+}^{0}(i\eps) G_{+}^{0}(i(\eps-\om)) G_{+}^{0}(i(\eps-\om+\om')) G_{+}^{0}(i(\eps + \om'))
\nonum \\
+&&{\xi_{k+q}^{(+)} - \xi_{k}^{(+)} - q \ov 2(i\om - q)}
~G_{+}^{0}(i\eps) G_{+}^{0}(i(\eps+\om)) G_{+}^{0}(i(\eps+\om+\om')) G_{+}^{0}(i(\eps + \om'))~].
\nonum
\eea
In the case of a quadratic spectrum ($\xi_k^{(+)} = vk + k^2/2m_+$ where we explicitly introduce the right movers mass $m_+$):
\bea
&&\Pi_{R}^{(2)}(i\om,q) = {q \ov m_+}~{1 \ov \beta^2} \sum_{\eps,\om'} \int {{dk dq'\ov (2\pi)^2}}~|V[q']|^2~\Pi_-^0(i\om',q')~[
\nonum \\
&&{2k \ov i\om - q}~~G_{+}^{0}(i\eps) G_{+}^{0}(i(\eps-\om)) G_{+}^{0}(i(\eps+\om)) G_{+}^{0}(i(\eps+\om'))
- q \left [ G_{+}^{0}(i\eps) \right ]^2 G_{+}^{0}(i(\eps-\om)) G_{+}^{0}(i(\eps+\om)) G_{+}^{0}(i(\eps+\om'))
\nonum \\
&&+ {2k-q \ov 4(i\om - q)}~G_{+}^{0}(i\eps) G_{+}^{0}(i(\eps-\om)) G_{+}^{0}(i(\eps-\om+\om')) G_{+}^{0}(i(\eps + \om'))
\nonum \\
&&+ {2k+q \ov 4(i\om - q)}~G_{+}^{0}(i\eps) G_{+}^{0}(i(\eps+\om)) G_{+}^{0}(i(\eps+\om+\om')) G_{+}^{0}(i(\eps + \om'))~].
\nonum
\eea
These four terms give a non-zero contribution only in the case of a non-linear spectrum. Notice that this is valid for all temperatures and for all interaction ranges (except for a change of sign which does not affect the interaction the variables $\om'$ and $q'$ have not been shifted during the splitting procedure). Further use of the splitting procedure shows that the lowest non-zero order is in $1/m^2$. The corresponding expression for the sum of the three diagrams is:
\bea
&&\Pi_{R}^{(2)}(i\om,q) =
\nonum \\
&&{1 \ov (i\om - q)^2}~{q^2 \ov m_+}~{1 \ov \beta^2} \sum_{\eps,\om'} \int {{dk dq'\ov (2\pi)^2}}~|V[q']|^2~\Pi_-^0(i\om',q') [G_{+}^{0}(i\eps)]^2 G_{+}^{0}(i(\eps+\om'))
\nonum \\
+&& {1 \ov (i\om - q)^2}~{q^2 \ov m_+^2}~{1 \ov \beta^2} \sum_{\eps,\om'} \int {{dk dq'\ov (2\pi)^2}}~|V[q']|^2~\Pi_-^0(i\om',q')~[
\nonum \\
+&& 2k^2~G_{+}^{0}(i\eps) G_{+}^{0}(i(\eps-\om)) G_{+}^{0}(i(\eps+\om)) G_{+}^{0}(i(\eps+\om'))
\nonum \\
-&& kq~(i\om - q)~
\left [ G_{+}^{0}(i\eps) \right ]^2 G_{+}^{0}(i(\eps-\om)) G_{+}^{0}(i(\eps+\om)) G_{+}^{0}(i(\eps+\om'))
\nonum \\
-&&{[2k-q]q \ov 4}~
[G_{+}^{0}(i\eps)]^2 G_{+}^{0}(i(\eps-\om)) G_{+}^{0}(i(\eps + \om'))
\nonum \\
+&&{[2k+q]q \ov 4}~
[G_{+}^{0}(i\eps)]^2 G_{+}^{0}(i(\eps+\om)) G_{+}^{0}(i(\eps + \om'))
\nonum \\
+&&{[2k-q]~[2(k+q')-q] \ov 8}~
G_{+}^{0}(i\eps) G_{+}^{0}(i(\eps-\om)) G_{+}^{0}(i(\eps-\om+\om')) G_{+}^{0}(i(\eps + \om'))
\nonum \\
+&&{[2k-q]~[2(k+q')-q] \ov 8}~G_{+}^{0}(i\eps) G_{+}^{0}(i(\eps+\om)) G_{+}^{0}(i(\eps+\om+\om')) G_{+}^{0}(i(\eps + \om'))~].
\nonum
\eea
The above equations are valid to all orders in $1/m$ and show that the first order correction is in $1/m^2$ (the first term is a pure mass-shell singularity and therefore vanishes in the limit $w \not=q$ we are interested in). The fact that $1/m^2$ is the first relevant order agrees with the corresponding bosonic cubic field theory, see {\it e.g.}~[\onlinecite{Samokhin}]. Moreover the generation of $1/m$ factors from our splitting procedure is due only to a single spicy of fermions, i.e. re-arrangement of Green's functions of right movers for the above R-diagrams (left movers for the corresponding L-diagrams). The mass appearing in the above expansion is therefore the one of right-movers, $m \equiv m_+$ (similarly the mass appearing in L-diagrams is the one of left-movers, $m \equiv m_-$).

To proceed further from the above equation, the easiest task is to compute the leading order in $1/m$ of the sum of R-diagrams. As has been discussed in Sec.~\ref{diagrammatics_results}, this amounts to focus on the tails of the density-density correlation function. For this purpose, we take the Green's functions in factor of the $1/m^2$ terms as the massless ones. Notice that this implies that the tails of R-diagrams (resp. L-diagrams) correspond to the scattering or a curved right-mover (resp. left-mover) density fluctuation on a linearized left-mover (resp. right-mover) density fluctuation. The final expression for the imaginary part of the polarization operator is:
\bea
&&\Im \Pi_R^{(2)R}(\om,q) = {-1 \ov 32 \pi v^2 (\om - vq)^3}~\left({q \ov m_+}\right)^2~|V({\om-vq \ov 2})|^2~\int {{dk dk'}}~[
\nonum \\
&&(2k-q)~[ n_F(vk-vq)-n_F(vk-vq+\om)]~[ n_F(-vk')-n_F(-vk'+{\om-vq \ov 2}) ]
~[ n_B(-{\om - vq \ov 2})+ n_F(vk+{\om-vq \ov 2})]
\nonum \\
&&+(2k+q)~[ n_F(vk+vq)-n_F(vk+vq-\om)]~[ n_F(-vk')-n_F(-vk'-{\om-vq \ov 2})]
~[ n_B({\om - vq \ov 2})+ n_F(vk-{\om-vq \ov 2})]~],
\nonum \\
&&\Im \Pi_L^{(2)R}(\om,q) = {+1 \ov 32 \pi v^2 (\om + vq)^3}~\left({q \ov m_-}\right)^2~|V({\om+vq \ov 2})|^2~\int {{dk dk'}}~[
\nonum \\
&&(2k-q)~[ n_F(-vk+vq)-n_F(-vk+vq+\om)]~[ n_F(vk')-n_F(vk'+{\om+vq \ov 2}) ]
~[ n_B(-{\om + vq \ov 2})+ n_F(-vk+{\om+vq \ov 2})]
\nonum \\
&&+(2k+q)~[ n_F(-vk-vq)-n_F(-vk-vq-\om)]~[ n_F(vk')-n_F(vk'-{\om+vq \ov 2})]
~[ n_B({\om + vq \ov 2})+ n_F(-vk-{\om+vq \ov 2})]~],
\nonum
\eea
were we included both the contributions of R diagrams ($\Im \Pi_R^{(2)R}(\om,q)$) and
L diagrams ($\Im \Pi_L^{(2)R}(\om,q)$). These are our expressions for the tails of the sum of R and L diagrams, in the lowest order in $1/m$, in second order in interactions, for any interaction range and at all temperatures. Notice that the integrals over wave-numbers are converging and that the overall expressions have the correct symmetry. Integrating over momentum yields Eqs.~(\ref{R}) and~(\ref{L}). These were further discussed in Sec.~\ref{diagrammatics_results}.

\subsection{Mixed (M1 and M2) diagrams with curvature}

We focus now on mixed diagrams, i.e. diagrams M1 and M2, with curvature. As for the chiral diagrams we consider the combination M1 + M2. Proceeding along the arguments for the vanishing of such combination in the massless case we generate a systematic expansion in $1/m$.

The expressions of the mixed diagrams are:
\bea
\Pi_{M1}^{(2)}(i\om,q) =&& {1 \ov \beta^3} \sum_{\eps_1,\eps_2,\om'} \int {{dk_1 dk_2 dq' \ov (2\pi)^3}}~V[q'] V[-q-q']
\nonum \\
&& G_{+}^{0}(i\eps_1,k_1) G_{+}^{0}(i(\eps_1+\om'),k_1+q') G_{+}^{0}(i(\eps_1-\om),k_1-q)
\nonum \\
&& G_{-}^{0}(i\eps_2,k_2) G_{-}^{0}(i(\eps_2+\om'),k_2+q') G_{-}^{0}(i(\eps_2-\om),k_2-q),
\nonum \\
\Pi_{M2}^{(2)}(i\om,q) =&& {1 \ov \beta^3} \sum_{\eps_1,\eps_2,\om'} \int {{dk_1 dk_2 dq' \ov (2\pi)^3}}~V[q'] V[-q-q']
\nonum \\
&& G_{+}^{0}(i\eps_1,k_1) G_{+}^{0}(i(\eps_1 - \om - \om'),k_1 - q - q') G_{+}^{0}(i(\eps_1-\om),k_1-q)
\nonum \\
&& G_{-}^{0}(i\eps_2,k_2) G_{-}^{0}(i(\eps_2+\om'),k_2+q') G_{-}^{0}(i(\eps_2-\om),k_2-q),
\nonum
\eea
where $V[q]$ is the interaction potential and the bare Green's function are curved. The triple integrals over momentum show that these are $3-$loop contributions.

Performing the shifts: $\eps_1 \rightarrow \eps_1 + \om + \om'$
and $k_1 \rightarrow k_1 + q + q'$ in M2, adding M1 and M2 with the left three-legged loop factorized yields:
\bea
\Pi_{M}^{(2)}(i\om,q) =&& 2\Pi_{M1}^{(2)}(i\om,q) + 2\Pi_{M2}^{(2)}(i\om,q) = {2 \ov \beta^3} \sum_{\eps_1,\eps_2,\om'} \int {{dk_1 dk_2 dq' \ov (2\pi)^3}}~V[q'] V[-q-q']
\nonum \\
&& G_{-}^{0}(i\eps_2,k_2) G_{-}^{0}(i(\eps_2+\om'),k_2+q') G_{-}^{0}(i(\eps_2-\om),k_2-q)
\nonum \\
&&[ G_{+}^{0}(i\eps_1,k_1) G_{+}^{0}(i(\eps_1+\om'),k_1+q') G_{+}^{0}(i(\eps_1-\om),k_1-q)
\nonum \\
&&+ G_{+}^{0}(i\eps_1,k_1) G_{+}^{0}(i(\eps_1 + \om + \om'),k_1 + q + q') G_{+}^{0}(i(\eps_1+\om'),k_1+q')],
\nonum
\eea
where the factor of $2$ has been taken into account (M1 and M2 appear twice).
We implement the splitting procedure by setting:
\bea
G_{+}^{0}(i\eps_1) G_{+}^{0}(i(\eps_1-\om)) =&&
{1 \ov i\om - vq} [ G_{+}^{0}(i(\eps_1-\om) - G_{+}^{0}(i(\eps_1)
- \left( {q^2 \ov 2 m_+} - {k_1 q \ov m_+} \right)G_{+}^{0}(i\eps_1) G_{+}^{0}(i(\eps_1-\om))],
\nonum
\eea
and:
\bea
G_{+}^{0}(i(\eps_1+\om')) G_{+}^{0}(i(\eps_1+\om'+\om)) =&&
{1 \ov i\om - vq} [ G_{+}^{0}(i(\eps_1+\om')) - G_{+}^{0}(i(\eps_1+\om'+\om))
\nonum \\
&&+ \left( {q^2 \ov 2 m_+} + {(k_1+q') q \ov m_+} \right)G_{+}^{0}(i(\eps_1+\om')) G_{+}^{0}(i(\eps_1+\om'+\om))],
\nonum
\eea
where one may recover initial notations by: $G_{+}^{0}(i\eps) \equiv G_{+}^{0}(i\eps,k)$.
This yields:
\bea
\Pi_{M}^{(2)}(i\om,q) =&& {1 \ov i\om - vq}{1 \ov \beta^3} \sum_{\eps_1,\eps_2,\om'} \int {{dk_1 dk_2 dq' \ov (2\pi)^3}}~V[q'] V[-q-q']
\nonum \\
&& G_{-}^{0}(i\eps_2) G_{-}^{0}(i(\eps_2+\om')) G_{-}^{0}(i(\eps_2-\om))
\nonum \\
&&[ - \left( {q^2 \ov 2 m_+} - {k_1 q \ov m_+} \right) G_{+}^{0}(i\eps_1) G_{+}^{0}(i(\eps_1-\om))G_{+}^{0}(i(\eps_1+\om'))
\nonum \\
&&+ \left( {q^2 \ov 2 m_+} + {(k_1+q') q \ov m_+} \right) G_{+}^{0}(i\eps_1) G_{+}^{0}(i(\eps_1+\om')) G_{+}^{0}(i(\eps_1+\om'+\om))],
\nonum
\eea
whereby we recover that the combination vanishes in the massless limit. Re-implementing the procedure to the last expression and after some lengthy algebraic manipulations the sum of M diagrams reads:
\bea
&&\Pi_{M}^{(2)}(i\om,q) = {1 \ov (i\om - vq)^2(i\om + vq)}~{q^2 \ov 2 m_+ m_-}~{1 \ov \beta^3} \sum_{\eps_1,\eps_2,\om'} \int {{dk_1 dk_2 dq' \ov (2\pi)^3}}~[
\nonum \\
&& - V[q'] V[q'+q] [2k_1-q][2k_2-q] G_{+}^{0}(i\eps_1) G_{+}^{0}(i(\eps_1+\om')) G_{-}^{0}(i\eps_2) G_{-}^{0}(i(\eps_2-\om)) G_{-}^{0}(i(\eps_2+\om'))
\nonum \\
&& + V[q'] V[q'-q] [2k_1+q][2k_2+q] G_{+}^{0}(i\eps_1) G_{+}^{0}(i(\eps_1+\om')) G_{-}^{0}(i\eps_2) G_{-}^{0}(i(\eps_2+\om)) G_{-}^{0}(i(\eps_2+\om'))
\nonum \\
&& - V[q'] V[q'+q] [2k_1-q][2k_2+q] G_{+}^{0}(i\eps_1) G_{+}^{0}(i(\eps_1+\om')) G_{-}^{0}(i\eps_2) G_{-}^{0}(i(\eps_2+\om)) G_{-}^{0}(i(\eps_2-\om'))
\nonum \\
&& + V[q'] V[q'-q] [2k_1+q][2k_2-q] G_{+}^{0}(i\eps_1) G_{+}^{0}(i(\eps_1+\om')) G_{-}^{0}(i\eps_2) G_{-}^{0}(i(\eps_2-\om)) G_{-}^{0}(i(\eps_2-\om'))].
\nonum
\eea
This expression satisfies the basic symmetries of the Matsubara polarization operator. Note that it is exact to all orders in $1/m$ as the bare Green's functions sitting in the integrand are curved. It shows that the mixed diagrams have an expansion starting as $1/m_+m_-$ so that curvature is crucial for both right- and left-movers, even for the tails. This is the root of our terminology (mixed) for such diagrams and a basic difference between them and the so-called chiral contributions R and L.

The tails are then easily derived by taking the Green's functions as the massless ones.
This also enables further re-arrangements of the last equation which reduces it to:
\bea
&&\Pi_{M}^{(2)}(i\om,q) = {1 \ov (i\om - vq)^2(i\om + vq)}~{q^2 \ov  m_+ m_-}~{1 \ov \beta^3} \sum_{\eps_1,\eps_2,\om'} \int {{dk_1 dk_2 dq' \ov (2\pi)^3}}~[
\nonum \\
&& - V[q'] V[q'+q] [2k_1-q][2k_2-q] G_{+}^{0}(i\eps_1) G_{+}^{0}(i(\eps_1+\om')) G_{-}^{0}(i\eps_2) G_{-}^{0}(i(\eps_2-\om)) G_{-}^{0}(i(\eps_2+\om'))
\nonum \\
&&  + V[q'] V[q'-q] [2k_1+q][2k_2+q] G_{+}^{0}(i\eps_1) G_{+}^{0}(i(\eps_1+\om')) G_{-}^{0}(i\eps_2) G_{-}^{0}(i(\eps_2+\om)) G_{-}^{0}(i(\eps_2+\om'))],
\nonum
\eea
where the Green's functions are those of fermions with linear spectrum.

Performing the integration over frequencies and going to the retarded function yields:
\bea
&&\Im \Pi_{M}^{(2)R}(\om,q) =
{1 \ov 8 \pi v (\om^2 - (vq)^2)^2}~{q^2 \ov  m_+ m_-}~V[{\om+vq \ov 2}] V[{\om-vq \ov 2}]
\int {{dk_1 dk_2}}~[
\nonum \\
&& + [2k_1-q][2k_2-q]~[n_F[vk_1]-n_F[vk_1-{\om+vq \ov 2}]]
~[n_F[-vk_2+vq]-n_F[-vk_2+vq+\om]]
\nonum \\
&& [n_B[-{\om+vq \ov 2}]+n_F[-vk_2+{\om+vq \ov 2}]]
\nonum \\
&& - [2k_1+q][2k_2+q]~[n_F[vk_1]-n_F[vk_1+{\om+vq \ov 2}]]
~[n_F[-vk_2-vq]-n_F[-vk_2-vq+\om]]~
\nonum \\
&& [n_B[+{\om+vq \ov 2}]+n_F[-vk_2-{\om+vq \ov 2}]]].
\nonum
\eea
Performing the momentum integrations yields Eq.~(\ref{M}) which was further discussed in Sec.~\ref{diagrammatics_results}.

\bs

\end{widetext}

\section{Conclusion}
\label{conclusion}

As a conclusion, we have developed a diagrammatic approach to the problem of the tails of the dynamic structure factor of 1D spinless fermions in clean quantum wires. Such tails correspond to the multi-pair excitation continuum of the 1D liquid. The main difficulty that we have managed to overcome was in rigorously taking into account of both interactions and band-curvature in the lowest meaningful order of perturbation theory: second order in interactions, {\it i.e.} $2-$pair excitation continuum, and second order in curvature.

Our methodology has revealed three classes of diagrams, see Fig.~\ref{Diagrams2nd}: two "chiral" (R and L) classes which bring divergent contributions in the limits $\om \rightarrow \pm vq$, {\it i.e.} near the single-pair excitation continuum, and a "mixed" (M) class (so-called Aslamasov-Larkin or Altshuler-Shklovskii type diagrams) which is crucial for the f-sum rule to be satisfied. These diagrams describe the lowest order non-trivial microscopic scattering mechanisms taking place within the multi-pair continuum, {\it i.e.} at the $2-$pair level. Such mechanisms are beyond the reach of the Tomonaga-Luttinger model or, equivalently, the corresponding diagrams appear as non-cancelable by the Dzyaloshinskii-Larkin theorem. This is summarized by Eqs.~(\ref{Results}) which display the expression of the high-frequency polarization-part for each class. The corresponding DSF is given by Eq.~(\ref{DSFtails}). The zero-temperature limit, Eq.~(\ref{DSF0T}), agrees with the results of Refs.~[\onlinecite{GlazmanDrag}] and~[\onlinecite{Affleck}]. At non-zero temperatures, the $2-$pair excitation contribution, Eq.~(\ref{DSFT}), dominates over the single-pair excitation contribution over the range of frequencies: $|q|T/k_F \ll \om \pm vq \ll T$ (substantial for: $q \ll k_F$). Our results are more general than the ones of Ref.~[\onlinecite{GlazmanDrag}] in the sense that we have added the effects of temperature and long-range interactions. In the same respect they appear complementary to the results of Ref.~[\onlinecite{Affleck}] which did not consider temperature effects but where interactions were taken into account to all orders.

These results further allowed us to compute some observables which may be of relevancy to experiments in the field (tunneling experiments, energy loss experiments, etc...), {\it e.g.} see [\onlinecite{experiments}]. In particular, we have shown that the $2-$pair excitation continuum gives rise to an optical fermion conductivity Eq.~(\ref{sigmatails}); this may be an interesting alternative source of conduction besides the proximity to a Mott insulating phase, {\it i.e.} Umklapp processes, considered in the literature [\onlinecite{GiamarchiOrganics}]. Moreover, pushing our perturbation theory to its limits we have estimated the attenuation rate of a coherent mode, $\Omega_q$, due to the $2-$pair excitation continuum, Eqs.~(\ref{damping}), (\ref{damping_0T}) and (\ref{damping_T}). Our results show a crossover from $\gamma_q \sim |q|^3$, cf. Eq.~(\ref{damping_0T}), at low temperatures to $\gamma_q \sim q^2$, cf. Eq.~(\ref{damping_T}), at high temperatures when the mode is acoustic, $\Omega_q \sim |q|$. The $q-$dependence of our $T=0$ result seems to agree with Ref.~[\onlinecite{PK}] -- even though these authors were closer to the singular line $vq$ than us -- but not with Ref.~[\onlinecite{Samokhin}]. It also seems to disagree with the most recent contributions: Refs.~[\onlinecite{Glazman2}] and [\onlinecite{Affleck}]. Finally, we have determined curvature corrections to the electron-electron scattering rate, Eq.~(\ref{scatt4}). This exercise has revealed interesting diagrams, the fourth-order diagrams in interaction among the fermions shown in Fig.~\ref{Diagrams2ndGF}, responsible for such corrections. Such diagrams contain some of the polarization parts of Fig.~\ref{Diagrams2nd} and therefore describe scattering mechanisms which are beyond the reach of the Tomonaga-Luttinger model. Their peculiarity is that they all contain a cut with 5 particle-lines. This implies that these fourth-order corrections correspond to three-body processes~\cite{Note:3body} among fermions in the clean quantum wire. From the tails of the polarization parts such corrections scale as: $T^3 / \eps_F^2$, and sublead the linear in $T$ Luttinger-liquid scattering rate.

The above results are all related to the tails of the DSF. In relation with the recent Refs.~[\onlinecite{PK,Glazman2,Affleck}] we leave it for future work to consider the singular limit, $\om \rightarrow v |q|$.

\acknowledgments

I am extremely grateful to the Abdus Salam ICTP for hospitality and providing a very rich working atmosphere. I would like to thank I.~Gornyi, I.~Lerner, A.~Mirlin, M.~Pustilnik and I.~Yurkevich for discussions and particularly V.~Kravtsov and O.~Yevtushenko with whom I have shared all progresses of the work, benefiting from their advice. I owe to V. Kravtsov the fact that I have been working on curvature problems; I am also indebted to L.~Glazman for pointing the problem of the DSF tails and sharing his knowledge on it and D.~Maslov for discussions on dephasing. Finally, I thank S. Brazovskii for his comments on the manuscript.


\begin{thebibliography}{99}

\bibitem{Bloch}
F.~Bloch, ZS. Phys. {\bf 81}, 363 (1933); Helv. Phys. Acta {\bf 7}, 385 (1934).

\bibitem{Tomonaga}
S.-I.~Tomonaga, Prog.~Theor.~Phys. {\bf 5}, 544 (1950).

\bibitem{Note:FS}
There is an additional assumption on the interaction processes: forward scattering, $g_2$ and $g_4$ processes.

\bibitem{Luttinger}
J. M. Luttinger, J. Math. Phys. {\bf 4} 1154 (1963).

\bibitem{DL}
I.~E.~Dzyaloshinskii and A.~I.~Larkin, Sov.~Phys.-JETP {\bf 38}, 202 (1974).

\bibitem{Bosonization}
A.~Luther and V.~J.~Peschel \prb {\bf 9}, 2911 (1974); S.~Coleman \prd {\bf 11}, 2088 (1975).

\bibitem{MonographeTG} T.~Giamarchi, "Quantum Physics in One Dimension", Oxford University Press, USA (2004).

\bibitem{MonographeAOG} A. O. Gogolin, A. A. Nersesyan and A. M. Tsvelik, "Bosonization and Strongly Correlated Systems", Cambridge University Press (2004).

\bibitem{BMN}
S.~Brazovskii, S.~Matveenko and P.~Nozi\`eres, J.~Phys.~I~France {\bf 4}, 571 (1994); S.~Brazovskii and S.~Matveenko Sov. Phys. JETP {\bf 78}, 892 (1994).

\bibitem{Kopietz}
P.~Kopietz and G.~Castilla, \prl {\bf 76}, 4777 (1996).

\bibitem{Samokhin}
K.~V.~Samokhin, J. Phys.: Cond. Matter {\bf 10}, L533 (1998).

\bibitem{GlazmanDrag}
M. Pustilnik, E.~G.~Mishchenko, L.~I.~Glazman and A.~V.~Andreev, \prl {\bf{91}}, 126805 (2003).

\bibitem{Wiegmann}
A.~G.~Abanov and P.~B.~Wiegmann, \prl {\bf 95}, 076402 (2005).

\bibitem{Rozhkov}
A.~V.~Rozhkov, Eur.~Phys.~J.~B {\bf 47}, 193 (2005).

\bibitem{PK}
P.~Pirooznia and P.~Kopietz, cond-mat/0512494.

\bibitem{Glazman2}
M. Pustilnik, M. Khodas, A. Kamenev and L.I. Glazman, cond-mat/0603458.

\bibitem{Affleck}
R. G. Pereira, J. Sirker, J.-S. Caux, R. Hagemans, J. M. Maillet, S. R. White and I. Affleck, cond-mat/0603681.

\bibitem{AGD}
A.~A.~Abrikosov, L.~P.~Gorkov and I.~E.~Dzyaloshinskii, "Methods of Quantum Field Theory in Statistical Physics", Dover Publics; Rev. English Edition (1977).

\bibitem{Nozieres}
D.~Pines and P. Nozi\`eres, "The Theory of Quantum Liquids", Vol. 1, 1989, Perseus Books Publishing.

\bibitem{BD}
S. Brazovskii and I.~Dzyaloshinskii Sov. Phys. JETP {\bf 44}, 1233 (1976).

\bibitem{g4_diagrams}
Including the $g_4-$process increases the number of diagrams and renders the diagrammatic approach more tedious. This process generates tails with the same functional form as the ones derived in this article. Only the numerical coefficient in Eqs.(\ref{Results}) would change. This is why we restrict ourselves to the minimal model with the $g_2-$process where less diagrams have to be computed.

\bibitem{DuBois}
D.~F.~DuBois, M.~G.~Kivelson, Phys. Rev. {\bf{186}}, 409 (1969).

\bibitem{Reizer1}
E.~G.~Mishchenko, M.~Yu.~Reizer and L.~I.~Glazman, \prb {\bf{69}}, 195302 (2004); M.~Yu.~Reizer and V.~M.~Vinokur, \prb {\bf{62}}, R16 306 (2000).

\bibitem{GiamarchiOrganics}
A. Schwartz, M. Dressel, G. Gr\"uner, V. Vescoli, L. Degiorgi and T. Giamarchi, \prb {\bf 58}, 1261 (1998); T. Giamarchi, S. Biermann, A. Georges and A. Lichtenstein, Proceedings of the ISCOM 2003 conference, Port-Bourgenay, Sept 2003 (cond-mat/0401268).

\bibitem{Note:plasmon_dispersion}
Notice that the velocity $v$ includes here the re-normalization by interactions, {\it i.e.} $v \equiv v~\sqrt{1 + V(q) / \pi v}$.

\bibitem{Maslov}
D.~Maslov in Les Houches Summer School LXXXI, Elsevier Science (2005), H. Bouchiat et al. Editors.

\bibitem{Note:FL}
Recall that in a Fermi liquid: $-\Im \Sigma \propto \eps^2 / \eps_F \ll \eps$.

\bibitem{experiments}
T. Pichler, M. Knupfer, M.S. Golden, J. Fink, A. Rinzler and R.E. Smalley, \prl {\bf 80}, 4729 (1998);
F. Perez, B. Jusserand and B. Etienne, Phys. Rev. B {\bf 60}, 13310 (1999).

\bibitem{Note:3body}
This follows from the fact that we could show, with the help of the Keldysh technique, that the fourth order self-energy diagrams of Fig.~\ref{Diagrams2ndGF} give rise to a $6-$fermion-occupation-function structure to the quantum kinetic equation, on-shell. Recently, relaxation phenomena in disordered wires were shown to require 3 bodies (2 electrons and 1 impurity), see [\onlinecite{GMP}].

\bibitem{GMP}
I.~V.~Gornyi, A.~D.~Mirlin and D.~G.~Polyakov, \prl {\bf 95}, 046404 (2005).

\end{thebibliography}
\end{document}